\documentclass[11pt]{article}
\usepackage{a4wide}
\usepackage{amsmath, amssymb, amscd, amsthm, amsfonts}
\usepackage[utf8]{inputenc}
\usepackage{graphicx}
\usepackage{hyperref}
\usepackage{cite}
\usepackage[dvipsnames]{xcolor}
\usepackage{caption}
\usepackage{subcaption}

\newcommand{\Tr}{{\rm Tr}}

\newcommand{\be}{\begin{equation}}
\newcommand{\ee}{\end{equation}}
\newcommand{\ba}{\begin{align}}
\newcommand{\ea}{\end{align}}

\begin{document}
{\renewcommand{\thefootnote}{\fnsymbol{footnote}}
		
\begin{center}
{\LARGE Effective mass and symmetry breaking in the IKKT matrix model from compactification} 
\vspace{1.5em}

Samuel Laliberte\footnote{e-mail address: {\tt samuel.laliberte@mail.mcgill.ca}}
\\
\vspace{1.5em}
Department of Physics, McGill University, Montr\'{e}al, QC, H3A 2T8, Canada\\

\vspace{1.5em}
\end{center}
}
	
\setcounter{footnote}{0}

\begin{abstract}
	\noindent The IKKT model is a promising candidate for a non-perturbative description of Type IIB superstring theory. It is known from analytic approaches and numerical simulations that the IKKT matrix model with a mass term admits interesting cosmological solutions. However, this mass term is often introduced by hand, and serves as a regulator in the theory. In the present paper, we show that an effective mass matrix can arise naturally in the IKKT model by imposing a toroidal compactification where the space-time fermions acquire anti-periodic boundary conditions. When six spatial dimensions are chosen to be compact, the effective mass matrix breaks the SO(1,9) space-time symmetry of the IKKT model to SO(1,3) $\times$ SO(6). This paves the way for space-time solutions of the IKKT model where SO(1,9) symmetry is naturally broken to SO(1,3) $\times$ SO(6).
\end{abstract}

\tableofcontents

\section{Introduction}

Superstring theory is a promising candidate for a self-consistent unified theory of quantum gravity. An interesting feature of the theory is that the dimensionality of space-time is not arbitrary, but comes from the consistency of the theory. Specifically, the theory is only consistently defined in ten space-time dimensions.  For this theory to describe our world, one must impose that six out of the nine spatial dimensions are compactified. This can be done in many ways, resulting in a vast landscape of effective descriptions of string theory in four dimensions. In addition to the four dimensional vacua, there exist other ways to consistently compactify string theory to an arbitrary number of dimensions, which results in vacua that are not four dimensional. Clarifying why four dimensional vacua are prefered in the theory remains an open question, which to this day does not have an answer in perturbative string theory.

Another area where perturbative string theory lacks predictive power is in the context of cosmology. While string theory can be used to make predictions about the universe at late time, the cosmic stringularity is not resolved generally in perturbative string theory \cite{Lawrence:2002aj,Liu:2002kb,Horowitz:2002mw,Berkooz:2002je}.  Therefore, in order to study the very early universe, and explain the dimensionality of our world, we definitely a non-perturbative description. 

There have been many proposals for a non-perbative descriptions of string theory, most of them relying on matrix models \cite{Banks:1996vh,Berenstein:2002jq,Ishibashi:1996xs}. Among these theories, the IKKT model \cite{Ishibashi:1996xs}, a non-perturbative description of Type IIB superstring theory, stands out as a natural choice to explain the birth of the universe. This model is built around the action
\be
S_{IKKT} = - \frac{1}{4g^2} \Tr [A^M , A^N]^2 - \frac{1}{2g^2} \Tr  \bar{\psi} \Gamma^M [A_M , \psi] \, ,
\label{eq:IKKT_action}
\ee
where large bosonic matrices $A^M$'s encode information about space-time, and large fermionic matrices $\psi$'s are added to preserve supersymmetry. In the action above, $g$ is a gauge coupling which is related to the string scale $l_s$ via $g \sim l_s^2$, and the indices are contracted using the minkowski metric in the mostly minus sign convension $\eta_{MN} = \text{diag}(1,-1,-1,...,-1)$. Given the causal structure of the space, $A^0$ encodes information about time and $A^i$ encodes information about space, where $i \in \{1,...,9\}$ labels the nine space dimensions.

Over the years, there have been many attempts to find solutions of the IKKT model that correspond to an emergent four-dimensional universe. The first steps towards finding these solutions were done by analysing the model using Gaussian expansion method \cite{Oda:2000im,Sugino:2001fn} to investigate symmetry breaking in the theory. Using this method, it was shown that the SO(10) symmetry of the Euclidean version of this model can be spontaneously broken to SO(4) \cite{Nishimura:2001sx,Kawai:2002jk,Aoyama:2006rk,Nishimura:2011xy}. Monte Carlo simulations have also shown consistant results \cite{Anagnostopoulos:2017gos,Anagnostopoulos:2020xai}, and a recent analysis in the context of the BFSS model has also shown progress in this direction \cite{Brahma:2022ifx}. Then, Monte-Carlo simulations of the Lorentzian model showed that an expanding (1+3)-dimensional space-time can emerge from an SO(1,9) symmetric state of the model after a critical time \cite{Nishimura:2012cb,Ito:2013qga,Ito:2015mxa,Ito:2015mem}. To achieve this result, approximations were made to avoid the sign problem of the Lorentzian theory. However, further studies have shown that these approximations which are no longer valid when the four space-time dimensions emerge \cite{Aoki:2019tby}. Since then, the Lorenzian model has been studied without this approximation using the Complex Langevin Method, where the emergence of four space-time dimensions remain a topic of study \cite{Hirasawa:2021xeh,Hatakeyama:2021ake,Hatakeyama:2022ybs,Anagnostopoulos:2022dak,Hirasawa:2022qzg,Hirasawa:2023lpb}.

Despite the challenges encountered in studying the IKKT model from a numerical point of view, it was found in \cite{Nishimura:2012cb} that an important feature seems to be required to obtain the emergence of (1+3)-dimensions. When (1+3)-dimensions become large, certain bounds used to regulate theory, 
\begin{align}
\frac{1}{N} \Tr(A_0)^2 & \leq \kappa \frac{1}{N} \Tr (A_i)^2 \quad , \quad \\
\frac{1}{N} \Tr(A_i)^2 & \leq \kappa L^2 \, ,
\end{align}
become saturated. Saturating the constraints above is equivalent to adding the following piece to the IKKT action
\be
S_{const.} = \frac{\tilde{\lambda}}{2} \Tr \left(A_0^2 - \kappa L^2 \right) - \frac{\lambda}{2} \Tr \left(A_i^2 - L^2\right) \, ,
\label{eq:const.}
\ee
where $\lambda$ and $\tilde{\lambda}$ are Langrange multipliers. Minimising the IKKT action in the presence of the constraint above, it was shown analytically \cite{Kim:2012mw} and numerically \cite{Hatakeyama:2019jyw} that various cosmological solutions of the equation of motions can be found. An important point to notice is that adding the constraint piece in equation \ref{eq:const.} to the IKKT action is equivalent to adding a mass term to the theory, which may or may not be Lorentz invariant depending on the choice of $\lambda$ and $\tilde{\lambda}$.  Hence, adding a mass term to the theory can lead to interesting cosmological solutions. In fact, various analyses of the Lorentzian IKKT model with a mass term have been done before, in which case it was shown that cosmological solutions can also be found \cite{Steinacker:2017bhb,Sperling:2018xrm,Sperling:2019xar,Karczmarek:2022ejn}.\footnote{See \cite{Stern:2014aqa,Chaney:2015mfa,Chaney:2015ktw,Chaney:2016npa,Stern:2018wud} for other deformations of the IKKT model which admit cosmological solutions, and \cite{Battista:2022hqn,Battista:2022vvl,Battista:2023glw} for recent progress in the study of cosmological solutions in the IKKT model.} 

Since an effective mass term arises as a possible explanation for the emergence of cosmological solutions, it seems natural to ask what conditions are necessary for a mass term to naturally appear in the theory, and what causes the symmetry of space-time to break from SO(1,9) to SO(1,3) $\times$ SO(6). In the present paper, we explore this question by studying compactifications of the IKKT model. We find that if six spatial dimensions are compactified in a way that supersymmetry is broken, space-time fermions are quenched and the IKKT model action develops an effective mass matrix that breaks the SO(1,9) symmetry of the model to SO(1,3) $\times$ SO(6). This leads the way for solutions of the IKKT model where the SO(1,9) symmetry of space-time is naturally broken to SO(1,3) $\times$ SO(6).

\subsection{Outline}

To obtain the mass matrix, we will proceed as follows. In section \ref{sec:compact}, we will Wick rotate the Lorentzian IKKT model by imposing the change of variables $A^0 \rightarrow i A^0$ and $\Gamma^i \rightarrow i \Gamma^i$ to obtain the Euclidean IKKT model action 
\be
S_{IKKT} = - \frac{1}{4g^2} \Tr [A^M , A^N]^2 - \frac{i}{2g^2} \Tr  \bar{\psi} \Gamma^M [A_M , \psi] \, .
\label{eq:Euclidean_IKKT_action}
\ee
This transformation to Euclidean space will be done to simplify computations. Then, we will compactify this action on a six-dimensional torus where the space-time fermions $\psi$ acquire anti-periodic boundary conditions, hence breaking supersymmetry. As a result, the IKKT model action under compactification will become equivalent to a six-dimensional Yang-Mills theory with the following action
\be
S_C = \frac{1}{2g_{eff}^2} \int \frac{d\sigma^6}{V} \Tr \left( \frac{1}{2} F_{a b}F^{a b} + D_a A_\mu D^a A^\mu - \frac{1}{2}[A^\mu , A^\nu]^2 +  \bar{\psi} \Gamma^{a} D_a \psi - i \bar{\psi} \Gamma^\mu [A_\mu , \psi]\right) \, ,
\ee
where we have substituted the mode expansion
\be
A^M = \sum_{n^a \in \mathbb{Z}^6 } A^M (n^a) e^{i n^a \sigma^a} \quad , \quad \psi = \sum_{r^a \in \mathbb{Z}^6 + 1/2} \psi(r^a) e^{i r^a \sigma^a} \, .
\ee
Here, $\mu \in \{0,...,3\}$ labels the non-compact directions, $a \in \{4,5,..,9\}$ labels the compact directions, $V =(2 \pi L)^6$ is the volume of the internal space, $g_{eff}^2 = g^2/N$ is an effective gauge coupling, and $N$ is a large integer that we will introduce later. In this six-dimensional Yang-Mills theory, the zero modes describe non-compact degrees of freedom, and the non-zero models describe interactions between these non-compact degrees of freedom. Hence, integrating out the non-zero modes in the theory, one can obtain a Wilsonian effective action for the non-compact degrees of freedom in the theory. In section \ref{sec:wilson}, we will compute this Wilsonian effective action from the expression
\be
S^0_{eff} = - \ln \left( \sideset{}{'}\prod_{n^a \in \mathbb{Z}^6} \prod_{r^b \in \mathbb{Z}^6 + 1/2} \int \mathcal{D} A^M(n^a) \mathcal{D} \psi(r^b) \, e^{- S_E}\right) \, .
\label{eq:eff_act}
\ee
Here, $\sideset{}{'}\prod$ means that we are not integrating over the zero modes $n^a = 0$ of the theory. This computation will be done in the decompactification limit $L \gg g_{eff}^{1/2}$, where perturbation theory is valid and we expect to obtain a result close to the IKKT action without the compactification constraint (equation \ref{eq:IKKT_action}). Carrying out the computation to leading order in perturbation theory and Wick rotating back to Lorentzian space, we will find that the effective action takes the form
\begin{align}
	S^0_{eff} & = - \frac{1}{4g_{eff}^2} \Tr [A^M(0), A^N(0)]^2 + \frac{1}{2} M_{MN}^2 \Tr(A^M(0) A^N(0))^2 + ... \, ,
	\label{eq:result}
\end{align}
where the mass matrix
\be
M^2_{MN} = 
\begin{bmatrix}
	\eta_{\mu \nu} M_4^2 & 0 \\
	0 & \eta_{a b} M_6^2
\end{bmatrix}
\ee
arises as a first order correction which breaks SO(1,9) symmetry to SO(1,3) $\times$ SO(6). In the expression above, the masses $M_4^2$ and $M_6^2$ take the values
\begin{align}
	M_4^2 & = 16 \left(S_{F_1} - S_{B_1} \right) \frac{N M}{L^2} \, , \\
	M_6^2 & = \frac{32}{3} \left(S_{F_1} - S_{B_1} \right) \frac{N M}{L^2} \, ,
\end{align}
where the constants $S_{B_1}$ and $S_{F_1}$ are determined by the following sums
\be
S_{B_1} = \sideset{}{'}\sum_{n^a \in \mathbb{Z}^6} \frac{1}{(2\pi n^a)^2} \quad , \quad S_{F_1} = \sum_{r^a \in \mathbb{Z}^6 + 1/2} \frac{1}{(2\pi r^a)^2} \, .
\ee
The sums $S_{B_1}$ are $S_{F_1}$ are divergent in the large $n^a$ and $r^a$ limit. However, the difference between these two sums is finite and takes the value $S_{F_1} - S_{B_1} \approx 0.0397887$ when evaluated numerically.

The reason why we obtain equation \ref{eq:result} and not equation \ref{eq:IKKT_action} in the decompactification limit is because of broken supersymmetry. Since the fermions have anti-periodic boundary conditions, the fermionic zero modes are projected away in the mode expansion.  Hence, the fermionic sector does not enter the zero-mode effective action. We are left with the bosonic part of the IKKT action, and a mass matrix coming from integrating out interactions between the zero modes degrees of freedom in the theory. If supersymmetry is restored by imposing that fermions have periodic boundary conditions, $r^a$ becomes summed over $\mathbb{Z}^6$ instead of $\mathbb{Z}^6 + 1/2$ in the sum $S_{F_1}$. In this case, the masses $M_4^2$ and $M_6^2$ vanish since $S_{B_1} = S_{B_2}$, the fermions acquire a zero-mode term, and we obtain the IKKT model action (equation \ref{eq:IKKT_action}) with an effective gauge coupling $g_{eff}$.

\section{Compactification of the IKKT model}
\label{sec:compact}

Compactifying a matrix model presents a different challenge than compactifying a field theory. For one, there are no free parameters in the matrix model that we can choose to be compact. Hence, we must impose conditions on the matrices themselves. To overcome this challenge, we will make use of the method of mirror images, which was first brought forward by Washington Taylor in the context of D-brane mechanics \cite{Taylor:1996ik}. This method proved successful to explain graviton scattering under toroidal compactification of the BFSS model \cite{Berenstein:1997vm}, and has recently been used to explain three gravition amplitudes \cite{Herderschee:2023pza} and soft theorems \cite{Herderschee:2023bnc} in this same model.

This method builds on the fact that toroidal compactification is equivalent to duplicating a fundamental region of the target space an infinite number of time along said direction. For example, let us suppose we wish to compactify the real line $x \in \mathbb{R}$ on a circle $S^1$ of radius $R$. One option would be to confine the real line to an interval $x \in [0,2 \pi R[$ where we impose periodic boundary condition. Another would be to invoke the fact that periodic boundary conditions are equivalent to duplicating the interval $[0,2 \pi R[$ an infinite number of times along the real line. In other words, each point on the real line can be associated to a point a distance $x \rightarrow x + 2 \pi R$ away from this point. The mathematical term for this operation is called going to the universal cover of the circle.

\begin{figure}[h]
	\centering
	\begin{minipage}{.5\textwidth}
		\centering
		\begin{align*} 
			A^\mu & = 
			\begin{pmatrix}
				... & \color{red}{...} & \color{red}{...} & \color{red}{...} & \color{red}{...} \\
				\color{red}{...} & A^\mu(0)  & \color{red}{A^\mu(1)} & \color{red}{A^\mu(2)} & \color{red}{...} \\
				\color{red}{...} & \color{red}{A^\mu(-1)} & A^\mu(0) & \color{red}{A^\mu(1)} & \color{red}{...} \\
				\color{red}{...} & \color{red}{A^\mu(-2)} & \color{red}{A^\mu(-1)} & A^\mu(0) & \color{red}{...} \\
				\color{red}{...} & \color{red}{...} & \color{red}{...} & \color{red}{...} & ...
			\end{pmatrix} \\
			A^a & = 
			\begin{pmatrix}
				... & \color{red}... & \color{red}... & \color{red}... & \color{red}... \\
				\color{red}... & A^a(0) - 2\pi L & \color{red}A^a(1) & \color{red}A^a(2) & \color{red}... \\
				\color{red}... & \color{red}A^a(-1) & A^a(0) & \color{red}A^a(1) & \color{red}... \\
				\color{red}... & \color{red}A^a(-2) & \color{red}A^a(-1) & A^a(0) + 2\pi L & \color{red}... \\
				\color{red}... & \color{red}... & \color{red}... & \color{red}... & ...
			\end{pmatrix} \\
		\end{align*}
	\end{minipage}%
	\begin{minipage}{.5\textwidth}
		\centering
		\includegraphics[width = 7cm]{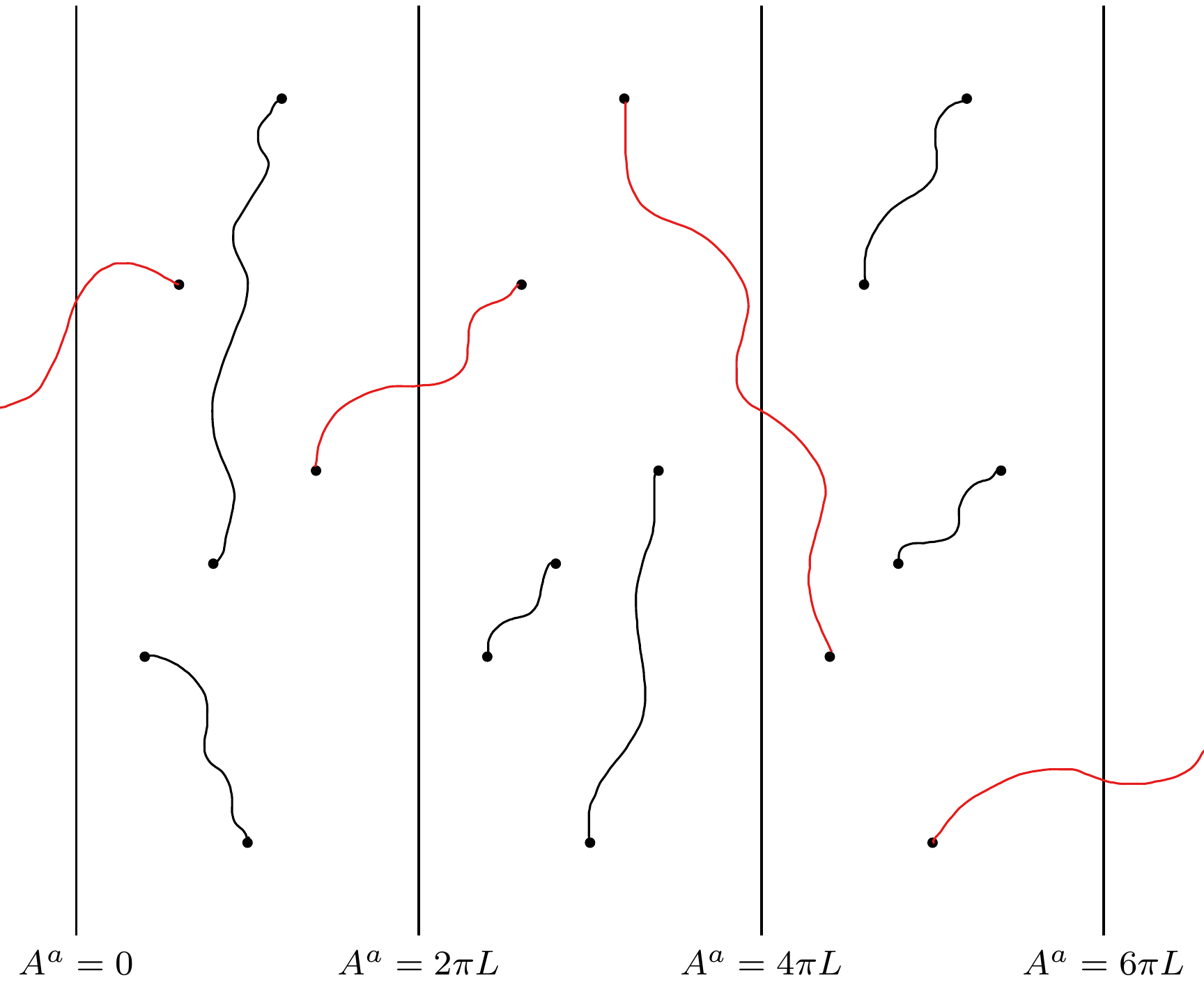}
	\end{minipage}
	\caption{Left: The diagonal blocks (black) describe the distribution of objects and their interactions in the duplicated regions, and the off-diagonal blocks (red) describe interactions between the duplicated regions. Right: Sketch of the duplicated regions along a compact direction $A^a$. The line between the black dots depict interactions inside (black) and across duplicated regions (red).}
	\label{fig:matrices}
\end{figure}

The same procedure can be applied to matrix models to impose a compactification. Since the matrix model describes a target space, we can impose that the target space contains duplicated objects in the direction we want to compactify in an attempt to replicate the effects of a compact space. To see how this is done in the context of the IKKT model, let us first Wick rotate the Lorentzian IKKT model to Euclidan space by imposing the change of variables $A^0 \rightarrow i A^0$ and $\Gamma^i \rightarrow i \Gamma^i$ in the Laurentzian IKKT model action. We obtain
\be
S_{IKKT} = - \frac{1}{4g^2} \Tr [A^M , A^N]^2 - \frac{i}{2g^2} \Tr  \bar{\psi} \Gamma^M [A_M , \psi] \, .
\ee
As previously mentioned, we will be interested in configurations of the IKKT model where six spatial directions $A^a$ are compact, and where fermions acquire anti-periodic boundary conditions. Such compactifications were first studied in the BFSS model \cite{Banks:1999tr}, and have more recently been used to obtain a thermal state of the IKKT model \cite{Laliberte:2023bai}. In the present case, we will generalise the approach taken in \cite{Laliberte:2023bai} to the case where six dimensions are compactified. To do this, we will invoke the existence of unitary operators $U^a$, which generate a translation in the $A^a$ direction of the target space. In addition, we impose that these operators commute with each other,
\be
U^a U^b = U^b U^a \, , 
\ee
so that translations in different compact directions can be made independently of each other. Following our previous discussion, capactifying the target space on a six-dimensional torus where fermions acquire anti-periodic boundary conditions should be equivalent to imposing the conditions
\begin{align}
	(U^b)^{-1} A^\mu U^b & = A^\mu \\
	(U^b)^{-1} A^a U^b & = A^a + 2 \pi L \delta_{ab}\\
	(U^b)^{-1} \psi U^b & = - \psi \, , \label{eq:cons_psi}
\end{align}
where $L$ is the torus radius.  Here, $\mu$ labels the non-compact space-time directions and $a$ labels the compact space directions. To solve the constraint equation above, we will use a approach similar to the one in \cite{Ganor:1996zk} and assume that the Hilbert space that the $A$'s and $\psi$'s act on has the tensor product form
\be
X = Y \otimes Z \, ,
\ee
where $X$ is a $M \times M$ matrix that will remain invariant under the translation, and $Z$ is a $N \times N$ matrix associated to the Hilbert space the translations act on. We will then invoke that $U^a$ takes the following form
\be
U^a = \mathbb{I}_M \otimes e^{-i 2\pi q^a} e^{-ip^a} \, ,
\label{eq:anz1}
\ee 
where $\mathbb{I}_M$ is the $M$-dimensional identity operator and $q^a$ and $p^b$ are operators that satisfy the Heisenberg algebra $[q^a , p^b] = i \delta_{ab}$.  With the form above, the unitary operator $U^a$ satisfies $(U^a)^{-1} q^a U^a = q^a + 1$, and generates a shift from $q^a$ to $q^a + 1$. The extra factor of $e^{-i 2\pi q^a}$ does not affect this shift. However, it will play a role in achieving the anti-periodic boundary conditions for the fermions. Next, we will note that a matrix of the form
\be
B = \sum_{n^a} B(n^a) \otimes e^{i n^a p^a} \, ,
\ee
satisfies $(U^a)^{-1} B U^a = B$ if $n^a$ is an integer, and $(U^a)^{-1} B U^a = - B$ if $n^a$ if a half-integer. Consequently, it's possible to solve the constraint equations by imposing that the matrices $A^\mu$, $A^a$, and $\psi$ take the following form
\begin{align}
	A^\mu & = \sum_{n^b \in \mathbb{Z}^6 } A^\mu(n^b) \otimes e^{i n^b p^b} \\
	A^a & = \sum_{n^b \in \mathbb{Z}^6 } A^a(n^b) \otimes e^{i n^b p^b} + 2 \pi L \, \mathbb{I}_M \otimes q^a \\
	\psi & =  \sum_{r^b \in \mathbb{Z}^6 + 1/2 } \psi(r^b) \otimes e^{i r^b p^b} \, .
\end{align}
In the expressions above, $n^b$ and $r^b$ are summed over $N$ integers and half-integers respectively, where $N$ is taken to be large but finite. It's possible to show that, when written in the $|q^a \rangle$ basis, that the matrices above take the block Toeplitz form depicted in Figure \ref{fig:matrices}. In this block Toeplitz form, the diagonal blocks describe the distribution of objects within an interval $[0,2 \pi L[$, and their interactions. The off-diagonal blocks, on their side, describe interactions between the duplicated fundamental regions.
Substituting the matrices above in the IKKT model action and using the identities
\be
[q^a , e^{inp^b}] = - n e^{inp^b} \delta_{ab} \quad , \quad \Tr e^{i (n \pm m) p^b} = N \delta(n \pm m) \, ,
\ee
we obtain the momentum space representation of the Yang-Mills action
\be
S_C = \frac{1}{2g_{eff}^2} \int \frac{d\sigma^6}{V} \Tr \left( \frac{1}{2} F_{a b}F^{a b} + D_a A_\mu D^a A^\mu - \frac{1}{2}[A^\mu , A^\nu]^2 +  \bar{\psi} \Gamma^{a} D_a \psi - i \bar{\psi} \Gamma^\mu [A_\mu , \psi]\right)
\label{eq:compact_IKKT_action}
\ee
where $g_{eff}^2 = g^2/N$ is an effective gauge coupling, $V =(2 \pi L)^6$ is the volume of the internal space and $A^M$ and $\psi$ are expanded using the mode decomposition
\be
A^M = \sum_{n^a \in \mathbb{Z}^6} A^M (n^a) \, e^{i n^a \sigma^a} \quad , \quad \psi = \sum_{r^a \in \mathbb{Z}^6 + 1/2} \psi(r^a) \, e^{i r^a \sigma^a} \, .
\ee
Here, $\sigma$ takes values inside the interval $[0,L^{-1}[$. Moreover, $A^M$ and $\psi$ respectively satisfy periodic and anti-periodic boundary conditions. In the mode expansion above, the zero modes are related to the distribution of objects and their interactions in the fundamental regions, and the non-zero modes are associated to interaction between fundamental regions \cite{Helling:2000bz}. In the decompactification limit $L \gg g_{eff}^{1/2}$, one should expect the fundamental regions to be far away from each other.  In the case, interactions will be suppressed, and we should obtain a theory which is approximately described by the dynamics of the zero-modes of the theory. This can be seen by looking at the mode expansion 
\begin{align*}
S_C & = - \frac{1}{4g_{eff}^2} \Tr [A^M(0) , A^N(0)]^2 +  \frac{1}{2g_{eff}^2} \sum_{n^a \in \mathbb{Z}^6} (2 \pi L n^a)^2 \Tr \left( A^M(-n^a) A^M(n^a) 
\right) \\
& \quad + \frac{1}{2g_{eff}^2} \sum_{r^a \in \mathbb{Z}^6 + 1/2} (2\pi L r^a i) \Tr \left( \bar{\psi}(r^a) \Gamma^{a} \psi(r^a) \right) + ... \, ,
\end{align*}
of the compact IKKT action. In the $L \gg g_{eff}^{1/2}$ limit, the non-zero winding modes $\omega_{n^a} = 2\pi L n^a$ and $\omega_{r^a} = 2\pi L r^a$ associated to the second and third term become heavy, and interactions become suppressed in the path integral. As a result, we expect the compact IKKT action to be effectively described by the zero modes of the system. This means we should recover the bosonic IKKT model action
\be
S_{C} = - \frac{1}{4g_{eff}^2} \Tr [A^M(0) , A^N(0)]^2 \, + ... \, ,
\label{eq:bosonic_IKKT}
\ee
and possible corrections coming from interactions between the fundamental regions. The fermions, in this case, do not contribute since their zero mode are projected away by the anti-periodic boundary conditions. As the radius of compactification $L$ decreases, one should expect that interactions become important, leading to more corrections to equation \ref{eq:bosonic_IKKT}. In the following sections, we will derive the leading corrections to equation \ref{eq:bosonic_IKKT} by evaluating a Wilsonian effective action for the zero modes $A^M(0)$ of the theory. In the limit where $L \gg g_{eff}^{1/2}$, we will see that the effective action of the zero modes aquires a mass matrix as first order correction, leading to symmetry breaking in the theory.

\section{Wilsonian effective action}
\label{sec:wilson}

To compute an effective action for the zero modes of the theory, which describes the non-compact degrees of freedom, we will adopt a Wilsonian approach. This appraoch will consist in integrating out the non-zero modes in the path integral in order to obtain an action that depends exclusively on the zero modes of the theory.

To see how this can be done, let us remind ourselves that a Wilsonian effective action can be used to find an effective description of the low energy modes of a theory by integrating out high energy modes above a cutoff $\Lambda$.  For example, let us consider the action $S[\Phi]$ associated to a scalar field $\Phi$.  To obtain the low energy effective action for some long wavelength modes $\Phi_L$, we can split the scalar field $\Phi = \Phi_L + \Phi_S$ into the contributions from $\Phi_L$ and the short wavelength component $\Phi_S$.  Then, the contribution of the short wavelength modes $\Phi_S$ can be integrated out in the partition function in the following way
\begin{align}
	Z & = \int \mathcal{D} \Phi e^{- S[\Phi]} \\
	& = \int \mathcal{D} \Phi_L \left( \int \mathcal{D} \Phi_S e^{S[\Phi_L + \Phi_S]}\right) \\
	& = \int \mathcal{D} \Phi_L e^{-S_{eff}[\Phi_L]} \, ,
\end{align}
to obtain a Wilsonian effective action $S_{eff}[\Phi_L]$ of the short wavelength component $\Phi_L$.  This Wilsonian effective action can be then computed from the expression
\be
S_{eff}[\Phi_L] = - \ln \left( \int \mathcal{D} \Phi_S e^{S[\Phi_L + \Phi_S]}\right) \, .
\ee
In the present case, we want to obtain an effective action of the zero modes $A^M(0)$ of the theory.  This means that, in the Wilsonian sense, we must integrate out all the non-zero modes $A^M(n^a)$ for $n^a \not= 0$ and $\psi(r^a)$ in the path integral.  To do this, we can split $A^M = A^M(0) + \sideset{}{'}\sum_{n^a \in \mathbb{Z}^6} A^M(n^a) e^{i n^a \sigma^a}$ into the zero-mode component $A^M(0)$ and and the non-zero-mode component $\sideset{}{'}\sum_{n^a \in \mathbb{Z}^6} A^M(n^a) e^{i n^a \sigma^a}$. Here, $\sideset{}{'}\sum$ means that we don't sum over the zero modes $n^a = 0$. We will then integrate out the non-zero-modes in the partition function in same way as for our scalar field example. For the compact IKKT action (equation \ref{eq:compact_IKKT_action}), this gives us
\begin{align}
	Z & = \prod_{n^a r^b \in \mathbb{Z}^6} \int \mathcal{D} A^M(n^a) \mathcal{D} \psi(r^b) \, e^{- S_C} \\
	& = \int \mathcal{D} A^M(0) \left( \sideset{}{'}\prod_{n^a \in \mathbb{Z}^6} \prod_{r^b \in \mathbb{Z}^6 + 1/2} \int \mathcal{D} A^M(n^a) \mathcal{D} \psi(r^b) \, e^{- S_C} \right) \\
	& = \int \mathcal{D} A^M(0) e^{- S^0_{eff}} \, ,
\end{align}
where 
\be
S^0_{eff} = - \ln \left( \sideset{}{'}\prod_{n^a \in \mathbb{Z}^6} \prod_{r^b \in \mathbb{Z}^6 + 1/2} \int \mathcal{D} A^M(n^a) \mathcal{D} \psi(r^b) \, e^{- S_C}\right) \, ,
\label{eq:eff_act}
\ee
can be identified as the zero-mode effective action.  Here again, we remind the reader that $\sideset{}{'}\prod$ means we integrate over all the modes $n^a \in \mathbb{Z}^6$ except the zero modes $n^a = 0$ of the theory. This means that $S^0_{eff}$ will depend exclusively on the zero modes $A^M(0)$ that haven't been integrated over. The goal of the next sections will be to compute the quantity above. This will be done using standard perturbative methods.

\subsection{Choice of gamma matrix representation and gauge fixing}

As a first step towards computing equation \ref{eq:eff_act}, we will choose a convenient representation for the gamma matrices that reflects the fact that $SO(10)$ symmetry is broken to $SO(4) \times SO(6)$ by our choice of compactification.  We will do this in a way to preserves the Majorana and Weyl conditions
\be
\Gamma_{11} \psi = \psi  \quad , \quad \bar{\psi} = \psi^T C_{10} 
\ee
which the fermions must satisfy for the theory to be supersymmetric.  Here, $\Gamma_{11}$ and $C_{10}$ are respectively the chirality operator and the charge conjugation operator in 10 dimensions.  In the present case, we will use the representation introduced in \cite{Yokoyama:2015yga} and consider Gamma matrices of the form 
\be
\Gamma^a = \tilde{\Gamma}^a \otimes 1 \quad , \quad \Gamma^\mu = \tilde{\Gamma}_7 \otimes \gamma^\mu \, ,
\ee
where $\tilde{\Gamma}^a$ are SO(6) gamma matrices, $\tilde{\Gamma}_7$ is the chirality operator these matrices and $\gamma^\mu$ are SO(4) gamma matrices (see \cite{Brink:1976bc} for other convenient representations).  We will further require that the SO(4) gamma matrices are in the Weyl representation 
\be
\gamma^\mu =
\begin{pmatrix}
	0 & \sigma^\mu \\
	\bar{\sigma}^\mu & 0 \,
\end{pmatrix}
\ee
where $\sigma^\mu$ and $\bar{\sigma}^\mu$ are Pauli 4-vectors which satisfy
\be
\bar{\sigma}_0 = \sigma_0 = 1 \quad , \quad \bar{\sigma}_i = - \sigma_i \quad , \quad \{\sigma_i , \sigma_j \} = - 2 \delta_{ij} \, .
\ee
In this representation, the chirality and charge conjugation operator for the 10 dimensional Gamma matrices take the form
\be
\Gamma_{11} = \tilde{\Gamma}_7 \otimes 
\begin{pmatrix}
	1 & 0 \\
	0 & -1
\end{pmatrix}
\quad , \quad
C_{10} = C_6 \otimes 
\begin{pmatrix}
	i \sigma_2 & 0 \\
	0 & - i \sigma_2
\end{pmatrix} \, .
\ee
Therefore, the Majorana and Weyl conditions reduce to
\be 
\psi = 
\begin{pmatrix}
	\psi^A_+ \\
	\psi^A_-
\end{pmatrix}
\quad , \quad 
\tilde{\Gamma}_7 \psi^A_\pm = \pm \psi^A_\pm 
\quad , \quad
\psi^A_\pm = \pm \epsilon^{AB} C_6 (\bar{\psi}^A_\pm)^T \, ,
\ee
where $A = 1 , 2$.  Given our choice of gamma matrices, the compact IKKT action takes the form
\begin{align}
	S_{C} & = \frac{1}{2g_{eff}^2} \int \frac{d\sigma^6}{V} \Tr \left( \frac{1}{2} F_{a b}F^{a b} + D_a A_\mu D^a A^\mu - \frac{1}{2}[A^\mu , A^\nu]^2 + \frac{1}{2}  \bar{\psi}^A_+ \tilde{\Gamma}^{a} \partial_a \psi^A_+ + \frac{1}{2}  \bar{\psi}^A_ - \tilde{\Gamma}^{a} \partial_a \psi^A_- \right. \\
	& \quad \left. - \frac{i}{2} \bar{\psi}^A_+ \tilde{\Gamma}^{a} [A_a ,  \psi^A_+] - \frac{i}{2} \bar{\psi}^A_- \tilde{\Gamma}^{a} [A_a ,  \psi^A_-] + \frac{i}{2} \bar{\psi}^A_+ (\sigma^\mu)^{AB} [A_\mu ,\psi^B_-] - \frac{i}{2} \bar{\psi}^A_- (\bar{\sigma}^\mu)^{AB} [A_\mu ,\psi^B_+] \right) \, .
\end{align}
In addition to our choice of gamma matrices, we will choose to work in the Lorenz gauge $\partial_a A^a = 0$.  This choice can be imposed by adding the ghost term 
\begin{align}
	S_{gh} = \frac{1}{g_{eff}^2} \int \frac{dx^{6}}{V} \Tr \left( \partial^a \bar{c} D_a c\right) \, .
\end{align}
to the compact IKKT action.

\subsection{Mode expansion}

Next, we will decompose the compact IKKT action into its different Fourier modes and separate the zero mode and the non-zero mode of the action. To do this, we will first separate the compact IKKT action $S_c = S_{kin} + S_{int}$ in a kinetic part
\begin{align}
	S_{kin} & = \int \frac{dx^6}{V} \Tr \left( \frac{1}{2} \partial_a A_N \partial^a A^N + \frac{1}{2}  \bar{\psi}^A_+ \tilde{\Gamma}^{a} \partial_a \psi^A_+ + \frac{1}{2}  \bar{\psi}^A_ - \tilde{\Gamma}^{a} \partial_a \psi^A_- + \partial_a \bar{c} \partial^a c \right) \, ,
\end{align}
and an interaction part
\begin{align}
	S_{int} & = \int \frac{dx^6}{V} \Tr \left( - i \partial_a A_N [A^a , A^N] - \frac{1}{4} [A^M , A^N]^2 - \frac{i}{2} \bar{\psi}^A_+ \tilde{\Gamma}^{a} [A_a ,  \psi^A_+] \right. \\
	& \left. - \frac{i}{2} \bar{\psi}^A_- \tilde{\Gamma}^{a} [A_a ,  \psi^A_-] + \frac{i}{2} \bar{\psi}^A_+ (\sigma^\mu)^{AB} [A_\mu ,\psi^B_-] - \frac{i}{2} \bar{\psi}^A_- (\bar{\sigma}^\mu)^{AB} [A_\mu ,\psi^B_+] - i \partial_{a} \bar{c} [A^a , c]  \right) \, .
\end{align}
Then, we will rescale the gauge fields, the fermions and the ghosts to make them dimensionless using the change of variable 
\be
A^M \rightarrow \lambda L A^M \quad , \quad \psi_+ \rightarrow \lambda L^{3/2} \psi_+ \quad , \quad \psi_- \rightarrow \lambda L^{3/2} \psi_- \quad , \quad c \rightarrow \lambda L c \, .
\ee
Here, $\lambda$ is a dimensionless parameter defined by $\lambda^2 \equiv g_{eff}^2/L^4$, which will play a role later in the perturbative expansion of equation \ref{eq:eff_act}. Finally, we will substitute the mode expansion 
\be
A^M = \sum_{n^a \in \mathbb{Z}^6} A^M(n^a) e^{i 2 \pi L n^a \sigma^a} \quad , \quad \psi = \sum_{r^a \in \mathbb{Z}^6 + 1/2} \psi(r^a) e^{i 2 \pi L r^a \sigma^a} \quad , \quad c = \sum_{n^a \in \mathbb{Z}^6} c(n^a) e^{i 2 \pi L n^a \sigma^a} \, ,
\ee
in $S_C = S_{kin} + S_{int}$. After this sustitution, the compact IKKT action can be written in the form $S_C = S_0 + S'_{kin} + S'_{int}$, where 
\be
S_0 = - \frac{\lambda^2}{4} \Tr [A^M(0), A^N(0)]^2 \,
\label{eq: S_0lambda}
\ee
is the zero mode part of the action, and
\begin{align}
	S'_{kin} & = \frac{1}{2} \sideset{}{'}\sum_{n^a \in \mathbb{Z}^6 } (2 \pi n^a)^2 \Tr \left( A_M(n^a) A^M(-n^a) \right) + \frac{1}{2} \sum_{r^a \in \mathbb{Z}^6 + 1/2 } (2\pi r^a i) \bar{\psi}^A_+(r^a) \tilde{\Gamma}^{a} \psi^A_+(r^a) \\
	& + \frac{1}{2} \sum_{r^a \in \mathbb{Z}^6 + 1/2 } (2\pi r^a i) \bar{\psi}^A_-(r^a) \tilde{\Gamma}^{a} \psi^A_-(r^a) + \sideset{}{'}\sum_{n^a \in \mathbb{Z}^6 } (2 \pi n^a)^2 \Tr \left( \bar{c}(n^a) c(n^a) \right) \, 
\end{align}
is the kinetic part where the zero modes, which do not contribute, are not summed over. The final term, corresponding to the interaction part where the zero modes have been removed, takes the form
\be
S'_{int} = \sum_{i = 1}^{5} V_i \, ,
\ee
where the $V_i$'s are given by
\begin{align}
	V_1 & = - \frac{\lambda^2}{4} \sideset{}{'}\sum_{n^a m^a l^a \in \mathbb{Z}^6 } \Tr \left( [A^M(-n^a - m^a - l^a) , A^N(n^a)][A_M(m^a) , A^N(l^a)] \right) \\
	\label{eq:V_1}
	V_2 & = - \lambda \sideset{}{'}\sum_{n^a m^a \in \mathbb{Z}^6} 2 \pi (n^a + m^a) \Tr \left( A_M(-n^a - m^a) [A^a(n^a) , A^M(m^a) ] \right) \\
	V_3 & = - \frac{i}{2} \lambda \sideset{}{'}\sum_{r^a \in \mathbb{Z}^6 + 1/2 \, , \, n^a \in \mathbb{Z}^6} \Tr \left( \bar{\psi}^A_+ (r^a + n^a) \tilde{\Gamma}^b [A_b(n^a) ,\psi^A_+(r^a)] \right) \\
	V_4 & = - \frac{i}{2} \lambda \sideset{}{'}\sum_{r^a \in \mathbb{Z}^6 + 1/2 \, , \, n^a \in \mathbb{Z}^6} \Tr \left( \bar{\psi}^A_- (r^a + n^a) \tilde{\Gamma}^b [A_b(n^a) ,\psi^A_-(r^a)] \right) \\
	V_5 & = \frac{i}{2} \lambda \sideset{}{'}\sum_{r^a \in \mathbb{Z}^6 + 1/2 \, , \, n^a \in \mathbb{Z}^6} \Tr \left( \bar{\psi}^A_+ (r^a + n^a) (\sigma^\mu)^{AB} [A_\mu(n^a) ,\psi^B_-(r^a)] \right) \\
	V_6 & = - \frac{i}{2} \lambda \sideset{}{'}\sum_{r^a \in \mathbb{Z}^6 + 1/2 \, , \, n^a \in \mathbb{Z}^6} \Tr \left( \bar{\psi}^A_- (r^a + n^a) (\bar{\sigma}^\mu)^{AB} [A_\mu(n^a) ,\psi^B_+(r^a)] \right) \\
	V_7 & = - \lambda \sideset{}{'}\sum_{n^a \in \mathbb{Z}^6 + 1/2 \, , \, m^a \in \mathbb{Z}^6} 2 \pi (n^a+m^b) \Tr \left( \bar{c}(n^a + m^a) [A_a(n^a) , c(m^a)] \right) \, .
	\label{eq:V_7}
\end{align}

\subsection{Zero mode effective action}

We are now in a position to evaluate the Wilsonian effective action for the zero modes of the theory. Before taking on the task of evaluating equation \ref{eq:eff_act}, let us pause and notice that the only free parameter in equation \ref{eq: S_0lambda} to \ref{eq:V_7} is the dimensionless quantity $\lambda$. In the computation that follows, $\lambda$ will play the role of expansion parameter. Since $S_0$ is an $\mathcal{O}(\lambda^2)$ quantity, we will only be concerned with corrections to $S_0$ that contribute at $\mathcal{O}(\lambda^2)$ order, neglecting the higher order corrections. This approximation is valid when $\lambda \ll 1$, or in other words when $L \gg g_{eff}^{1/2}$. In the IKKT model, $g^2$ is related to the string scale $l_s$ via $g^2 \sim l_s^4$. Hence, our approximation will be valid when the compactification radius $L$ is much larger than the string lenght $l_s$.

To evaluate \ref{eq:eff_act}, we will first substitute $S_E = S_0 + S'_{kin} + S'_{int}$ in our definition for the zero mode effective action of the theory (equation \ref{eq:eff_act}). We obtain
\begin{align*}
	S^0_{eff} & = - \ln \left( \sideset{}{'}\prod_{n^a \in \mathbb{Z}^6} \prod_{r^b \in \mathbb{Z}^6 + 1/2} \int \mathcal{D} A^M(n^a) \mathcal{D} \psi(r^b) \, e^{- S_C}\right) \\
	& = - \ln \left( e^{- S_0} \sideset{}{'}\prod_{n^a \in \mathbb{Z}^6} \prod_{r^b \in \mathbb{Z}^6 + 1/2} \int \mathcal{D} A^M(n^a) \mathcal{D} \psi(r^b) \, e^{- S'_{kin} - S'_{int}} \right) \\
	& = S_0 - \ln Z_{kin} - \ln \langle e^{- S'_{int}} \rangle \, ,
\end{align*}
where we have defined
\be
Z_{kin} = \sideset{}{'}\prod_{n^a \in \mathbb{Z}^6} \prod_{r^b \in \mathbb{Z}^6 + 1/2} \mathcal{D} A^M(n^a) \mathcal{D} \psi(r^b) \, e^{- S'_{kin}} \quad , \quad \langle \, . \, \rangle = \frac{1}{Z_{kin}} \sideset{}{'} \prod_{n^a \in \mathbb{Z}^6} \prod_{r^b \in \mathbb{Z}^6 +1/2} \int \mathcal{D} A^M(n^a) \mathcal{D} \psi(r^b) \, . \, e^{- S'_{kin}} \, .
\ee
As expected, the first term lets us recover the bosonic part of the IKKT action.  The second term, on its side, does not depend on $A^M(0)$ and is non-dynamical.  For this reason, we will simply ignore it.  Finally, we have the term $- \ln \langle e^{- S_{int}} \rangle$ which is dynamical and will bring correction to the bosonic IKKT action.  This term can be evaluated perturbatively by expanding it in the form 
\begin{align*}
	- \ln \langle e^{- S_{int}} \rangle & = \langle  S_{int} - \frac{1}{2} S_{int}^2 + ...\rangle_c \\
	& = \langle V_1 \rangle_c - \frac{1}{2} \langle V_2^2 \rangle_c - \frac{1}{2} \langle V_3^2 \rangle_c - \frac{1}{2} \langle V_4^2 \rangle_c - \langle V_5 V_6 \rangle_c - \frac{1}{2} \langle V_7^2 \rangle_c + ...
\end{align*}
where $\langle . \rangle_c$ denotes the fact that only connected diagrams contribute to the expectation value. In the expression above, we have only kept the terms that contribute to leading order ($\mathcal{O}(\lambda^2)$). All other contributions from the vertex terms (equation \ref{eq:V_1} to \ref{eq:V_7}) either vanish, contribute at next to leading ($\mathcal{O}(\lambda^4)$) order, or at a higher order in the expansion parameter $\lambda$. To evaluate the quantities above, it's useful to write down the two point functions
\begin{align}
	\langle A^M(n^a) A^N (m^a) \rangle & = \frac{ \delta_{MN} \delta_{n^a + m^a, 0}}{(2 \pi n^a)^2} \label{eq:two_point1} \\
	\langle \bar{\psi}^A_{+\alpha}(r^a) \psi^B_{+\beta}(s^a) \rangle & = - i \frac{ 2 \pi r^a  \tilde{\Gamma}^a_{\alpha \beta} \delta_{AB} \delta_{r^a, s^a}}{(2 \pi r^a)^2} \\
	\langle \bar{\psi}^A_{-\alpha}(r^a) \psi^B_{-\beta}(s^a) \rangle & = - i \frac{ 2 \pi r^a \tilde{\Gamma}^a_{\alpha \beta} \delta_{AB} \delta_{r^a,s^a}}{(2 \pi r^a)^2} \\
	\langle \bar{c}(n^a) c(m^a) \rangle & = \frac{ \delta_{n^a , m^b}}{(2 \pi n^a)^2} \, , \label{eq:two_point4}
\end{align}
for the gauge fields, the fermions and the ghosts.\footnote{In equation \ref{eq:two_point1} to \ref{eq:two_point4}, we did not write down the matrix indices to avoid cluttering the notation. Here, the two point functions of any two matrices $A_{ab}$ and $B_{cd}$ should take the form $\langle A_{ab} B_{cd} \rangle \sim \delta_{ad}\delta_{bc}$, where $a,b,c$ and $d$ are matrix indices.}  Using the two point functions above, we find 
\begin{align*}
	\langle V_1 \rangle_c & = 9 \lambda^2 M S_{B_1} \Tr (A^{N}(0))^2 \\
	\langle V_2^2 \rangle_c & = 2 \lambda^2 M \left( \left( 17 S_{B_2} + S_{B_1} \right) \Tr(A_0^{a})^2 + S_{B_1} \Tr(A^{\mu}(0))^2 \right)\\
	\langle V_3^2 \rangle_c & = - 8 \lambda^2 M (2 S_{F_2} - S_{F_1}) \Tr(A^{a}(0))^2 \\
	\langle V_4^2 \rangle_c & = - 8 \lambda^2 M (2 S_{F_2} - S_{F_1}) \Tr(A^{a}(0))^2 \\
	\langle V_5 V_6 \rangle_c & = 8 \lambda^2 M S_{F_1} \Tr(A^\mu(0))^2 \\
	\langle V_7^2 \rangle_c & = - 2 \lambda^2 M S_{B_2} \Tr(A^{a}(0))^2 
\end{align*}
where $S_{B_1}$ , $S_{B_2}$ , $S_{F_1}$ and $S_{F_2}$ are defined as follows
\be
S_{B_1} = \sideset{}{'}\sum_{n^a \in \mathbb{Z}^6} \frac{1}{(2\pi n^a)^2} \quad , \quad S_{F_1} = \sum_{r^a \in \mathbb{Z}^6 + 1/2} \frac{1}{(2\pi r^a)^2}
\ee
\be
S_{B_2} = \sideset{}{'}\sum_{n^a \in \mathbb{Z}^6} \frac{(2\pi n^1)^2}{(2\pi n^a)^4} \quad , \quad S_{F_2} = \sum_{r^a \in \mathbb{Z}^6 + 1/2} \frac{(2\pi r^1)^2}{(2\pi r^a)^4} \, .
\ee
Adding each term in the expansion, we find
\be
	- \ln \langle e^{i S_{int}} \rangle = - 8 \left( S_{F_1} - S_{B_1} \right) \lambda^2 M \Tr(A^\mu(0))^2 - 8 \left(S_{F_1} - S_{B_1} - 2 \left( S_{F_2} - S_{B_2} \right) \right) \lambda^2 M \Tr(A^a(0))^2 + \mathcal{O}(\lambda^4) \, .
\ee
The expression above can be simplified by noting that $S_{B_1} = 6 S_{B_2}$ and $S_{F_1} = 6 S_{F_2}$. Adding the corrections terms to $S_0$, the zero mode effective action at $\mathcal{O}(\lambda^2)$ takes the form
\begin{align}
	S^0_{eff} & = \left(- \frac{1}{4} \Tr [A^M(0), A^N(0)]^2 - 8 \left(S_{F_1} - S_{B_1} \right) M  \Tr(A^\mu(0))^2 - \frac{16}{3} \left(S_{F_1} - S_{B_1} \right) \Tr(A^a(0))^2 \right)\lambda^2 + \mathcal{O}(\lambda^4) \, .
\end{align}
Hence, we find that at leading order, the corrections to $S_0$ take the form of two mass terms: one associated to the non-compact directions $A^\mu$, and one associated to the compact directions $A^a$. As expected, these corrections break the $SO(10)$ symmetry of the target space to $SO(3) \times SO(6)$. This is to be expected since by making the choice to compactify six spatial dimensions, we are picking six special directions in space. The zero mode effective action at leading order in perturbation theory reflects this fact.

The expression above can be more neatly written after undoing our previous change of variable via $A^M \rightarrow \lambda^{-1} L^{-1} A^M$. In passing, we will also go back to Lorentzian signature by imposing $A^0 \rightarrow -i A^0$. In this case, the effective action takes the form
\begin{align}
	S^0_{eff} & = - \frac{1}{4g_{eff}^2} \Tr [A^M(0), A^N(0)]^2 + \frac{1}{2} M_{MN}^2 \Tr(A^M(0) A^N(0))^2 + ... \, ,
\end{align}
where we have defined a mass matrix
\be
M^2_{MN} = 
\begin{bmatrix}
	\eta_{\mu \nu} M_4^2 & 0 \\
	0 & \eta_{a b} M_6^2 \, ,
\end{bmatrix}
\ee
which includes two mass terms
\begin{align}
	M_4^2 & = 16 \left(S_{F_1} - S_{B_1} \right) \frac{N M}{L^2} \, , \\
	M_6^2 & = \frac{32}{3} \left(S_{F_1} - S_{B_1} \right) \frac{N M}{L^2} \, .
\end{align}
In the expression above, the sums $S_{B1}$ and $S_{F1}$ individually diverge in the limit where $N$ is large. However, it's possible to isolate the divergence in these sums by rewriting them as an integral and using Poisson resummation. What we find is rather interesting. It turns out that $S_{B1}$ and $S_{F_1}$ have the same divergent piece which is canceled by the difference $S_{F_1} - S_{B_1}$. It is then possible to evaluate difference numerically, which gives $S_{F_1} - S_{B_1} \approx 0.0397887$ . A detailed derivation of this result can be found in Appendix \ref{ap:Epstein}.

Notice that the breaking of supersymmetry plays a crucial role in obtaining non-vanishing masses $M_4^2$ and $M_6^2$. If supersymmetry is restored by imposing that fermions have periodic boundary conditions, then $r^a$ becomes summed over $\mathbb{Z}^6$ instead of $\mathbb{Z}^6 + 1/2$, and the masses vanish since $S_{B_1} = S_{B_2}$. Moreover, the fermions are indeed projected away by the anti-periodic boundary conditions, as expected. When suppersymmetry is restored, the fermions have zero modes terms that will appear at leading order in perturbation theory, and we recover the non-compact IKKT model action (equation \ref{eq:IKKT_action}) with an effective gauge coupling $g_{eff}$.

Moreover, notice that the mass term correction arise at leading order when integrating out the non-zero modes of the theory. This means that, in the decompactification limit, one cannot ignore residual interactions between duplicated regions. This potentially implies that interactions between regions are long ranged, and cannot be ignored even at large distances. 

A consequence of this phenomenon seems to be the breaking of gauge invariance in the fundamental regions. Since interactions between regions cannot be ignored, the theory develops an effective potential that takes the form of a mass term. This mass term, which impacts the distribution of objects and their interactions in the fundamental regions, also breaks the gauge invariance of the theory.\footnote{The IKKT model action is invariant under the gauge variations $\delta A^M = i [A^M , \alpha]$ and $\delta \psi = i [\psi , \alpha]$, where $\alpha$ is an arbitrary matrix. Including a mass term in the theory breaks this symmetry.} One may view this as being problematic since, naively, it should be expected that gauge invariance is preserved in the decompactification limit. This intuition comes from the fact that in the decompactification limit, we should recover the same theory we started with, along with the same symmetries. However, we should remind ourself that this is not the case when compactifying matrix theories. Instead of recovering the initial system, we recover a large $N$ number of copies of the initial system, as reflected by the overall factor of $N$ in the equation which is absorbed in the effective coupling $g_{eff}^2 = g^2 /N$. These copies come from the fact that we have duplicated a fundamental region $N$ times along the compact directions. Since we don't recover the same system we stated with, it's possible that some symmetries of the original system are not preserved. In the present case, we find that gauge symmetry in the fundamental regions is dependant on the structure of the interaction between them. If supersymmetry is preserved, interactions vanish and gauge symmetry is preserved. If supersymmetry is broken, the gauge symmetry is broken.

It is worth noting that compactifying a matrix theory on a higher dimensional torus can lead to some issues. For example, in the BFSS matrix model, decoupling breaks down when the theory is compactified on $T^k$ where $k > 5$ (see \cite{Banks:1999az} for more detail). However, this problem only arises when the compactification radius $L$ is taken to be small, and the system starts behaving like a dual quantum field theory. In the present case, the compactification radius is taken to be large, and the obtained system is closer to the IKKT model than a dual quantum field theory. Hence, we do not expect this issue to arise here. \footnote{We thank Savdeep Sethi for bringing this point to our attention.}

\section{Conclusion and discussion}

In this paper, we compactified the IKKT matrix model on a six-dimensional torus where the space-time fermions acquire anti-periodic boundary conditions, and we found that the Wilsonian effective action for the non-compact degrees of freedom in the theory acquires an effective mass term which breaks the SO(1,9) symmetry of the IKKT model to SO(1,3) $\times$ SO(6).  This mass matrix arises as a result of broken supersymmetry. If supersymmetry is restored, the conventional IKKT action (equation \ref{eq:IKKT_action}) is recovered. 

It would be interesting to see if the equations of motion of the effective action we have found have interesting cosmological solutions. Given that the SO(1,9) space-time symmetry of the IKKT model is broken to SO(1,3) $\times$ SO(6), one may expect there exist solutions where three space dimension expand, and the six other stay small. In this case, it might be possible that a SUSY breaking compactification is responsible for the emergence of three large space dimensions in recent numerical simulations of the IKKT model.

Assuming interesting cosmological solutions exist, it might be possible to use them to test recent predictions in matrix cosmology, one of them being the scale invariance of cosmological perturbations \cite{Laliberte:2023bai,Brahma:2021tkh} (see for \cite{Brahma:2022ikl} a summary of progress and challenges in these scenarios). Another avenue of research would be to test a recent space-time metric proposal in the IKKT matrix model \cite{Brahma:2022dsd} using these solutions, or repeat our anaylsis in the BFSS matrix model. In this case, one may find a possible connection with cosmological scenarios found in non-supersymmetric string theories \cite{Martinec:2009ks}.

Another exciting perspective is that higher order correction to the Wilsonian effective allow for Fuzzy de Sitter space solutions \cite{Buric:2015wta,Buric:2017yes,Buric:2019yau,Brkic:2022mcc}.  For example, Fuzzy dS$_4$ is described by four "Pauli-Lubanski" vectors that act as Casimir operators of the SO(1,4) group. Since these operators are built out of Lorentz generators of the SO(1,4) group, they satisfy well-known commutation relations. It would be interesting to see if these commutation relations are solutions of the IKKT model under compactification when higher order corrections are considered.

Finally, it is worth to mention that effective mass terms have been found in matrix models before, notably in the following work \cite{Mandal:2009vz,Mandal:2011hb}. However, in this case, the analysis was done for bosonic (1+D) and (2+D)-dimensional Yang-Mills theories where all all but one or two of the space-time matrices are integrated out. (1+D) and (2+D)-dimensional Yang-Mills theories can be viewed as a (1+D) and (2+D)-dimensional IKKT model where one or two dimensions are compactified on a torus. Hence, our analysis can be viewed as a special case of this work in which we consider a (6+4)-dimensional Yang-Mills theory where fermions are included, supersymmetry is broken, and all bosonic matrices are remain in the effective action. Contrary to \cite{Mandal:2009vz,Mandal:2011hb}, we have restrained ourselves to the limit where the compactification radius $L$ is large but finite. It would be interesting to see if phase transitions appear as we decrease the compactification radius.

\section*{Acknowledgements}

The author wishes to thank Robert Brandenberger, Keshav Dasgupta, Suddhasattwa Brahma and Savdeep Sethi for useful comments on the draft of this paper. Discussions with Herman Verlinde and Dionysios Anninos are also aknowledged. Additionally, the author wishes to thank the Erwin Schrodinger International Institute for Mathematics and Physics (ESI) for the hospitality during the workshop “Large-N Matrix Models and Emergent Geometry” in Vienna, where many good discussions with other participants helped shape the direction of this project. The author is supported in part by the Fonds de recherche du Québec (FRQNT). The research at McGill is supported in part by funds from NSERC and the Canada Research Chair program.

\appendix

\section{Epstein Series Regularisation}
\label{ap:Epstein}

When deriving the zero mode effective action of the IKKT model, we encountered the sums $S_{B1}$ and $S_{F1}$. These sums involve Epstein and Epstein-Hurwitz series that are divergent in the limit where $N \rightarrow \infty$.  In the present section, we show that the divergent part of these sums can be isolated by introducing a regulator in the sums. When this is done, we find that $S_{B1}$ and $S_{F1}$ have the same divergent part, which is canceled in the difference $S_{F1} - S_{B1}$. Evaluating the difference numerically, we find $S_{F1} - S_{B1} \approx  0.0397887$.

\subsection{Bosonic Sum}

We will first start by treating the divergence of the bosonic sum
\be
S_{B_1} = \sideset{}{'}\sum_{\vec{n} \in \mathbb{Z}^6} \frac{1}{|2\pi \vec{n}|^2} \, .
\ee
Here, we will use the vector notation $n^a = \vec{n}$ for simplicity. This sum involves the Epstein series
\be
E_B = \sideset{}{'}\sum_{\vec{n} \in \mathbb{Z}^d} \frac{1}{|\vec{n}|^2} \, ,
\ee
which diverges when $d/2 > 1$. To treat the divergence, we will modify the sum to include a UV regulator. Let us consider the expression
\be
\sideset{}{'}\sum_{\vec{n} \in \mathbb{Z}^d} \frac{e^{-\alpha^2 |\vec{n}|^2}}{|\vec{n}|^2} \, .
\label{eq:tamed_E_B}
\ee
Here, $\alpha^2$ plays the role of cutoff which truncates the modes above $N \sim \alpha^{-1}$ out of the sum, hence taming the divergence. In the limit where $\alpha^2 \rightarrow 0$, all modes contridute to the sum and the expression above reduces to $E_B$. Equation \ref{eq:tamed_E_B} can rewritten in integral form using the property
\be
\frac{1}{|\vec{n}|^2} = \int_0^{\infty} dt e^{-t |\vec{n}|^2} \, .
\label{eq:int_indentity1}
\ee
We obtain
\begin{align}
\sideset{}{'}\sum_{\vec{n} \in \mathbb{Z}^d} \frac{e^{-\alpha^2 |\vec{n}|^2}}{|\vec{n}|^2} & = \int_0^\infty dt \sideset{}{'}\sum_{\vec{n} \in \mathbb{Z}^d} e^{-(t + \alpha^2) |\vec{n}|^2} \\
& = \pi \int_{\alpha^2/\pi}^{\infty} dt \left( \theta^d(t) - 1 \right) \, ,
\label{eq:intB}
\end{align}
where we made use of the function
\be
\theta(t) = \sum_{n = -\infty}^{\infty} e^{-\pi t n^2} \, .
\ee
Since $\theta(t) \sim t^{-1/2}$ when $t \rightarrow 0$, the integrand in equation \ref{eq:intB} diverges in the limit when the regulator $\alpha^2$ goes to zero.  To deal with the divergent part of the integral, we will rewrite the part of the integral in the interval $t \in [\alpha^2/\pi , 1]$ by making use of the property
\be
\theta(t) = \frac{1}{t^{1/2}} \theta(1/t) \, ,
\ee
which can be derived using Poisson's resummation formula. Substituting the expression above in \ref{eq:intB}, we obtain
\be
\int_{\alpha^2/\pi}^{1} dt \left( \theta^d(t) - 1 \right) = \int_{1}^{\pi/\alpha^2} dt \, t^{d/2-1} \left( \theta^d(t) - 1 \right) - \frac{d}{d-2} + \alpha^2 + \frac{2}{d-2} \left( \frac{\pi}{\alpha^2} \right)^{d/2-1} \, .
\ee
In the expression above, the integral is finite for all values of $d$. Hence, for $d/2 > 1$, the only divergent piece when $\alpha^2 \rightarrow 0$ comes from the last term which is inversely propotional to $\alpha^2$. Piecing everything together and letting $\alpha^2$ go to zero, we obtain
\be 
E_B = \pi \left(\int_{1}^{\infty} dt \left(1 + t^{d/2-1} \right)\left( \theta^d(t) - 1 \right)  - \frac{d}{d-2} + \frac{2}{d-2} \left( \frac{\pi}{\alpha^2} \right)^{d/2-1} \right) \, .
\ee
When $d=6$, which is the case we are interested in, substituting the value of the $E_B$ in $S_{B_1}$ gives us
\be
S_{B_1} = \frac{1}{4\pi} \left(\int_{1}^{\infty} dt \left(1 + t^{2} \right)\left( \theta^6(t) - 1 \right)  - \frac{3}{2} + \frac{1}{2} \left( \frac{\pi}{\alpha^2} \right)^{2} \right) \, .
\ee

\subsection{Fermionic Sum}

Finally, we will evalutate the fermionic sum
\be
S_{F_1} = \sum_{\vec{n} \in \mathbb{Z}^6+1/2} \frac{1}{|2\pi \vec{n}|^2} \, ,
\label{eq:SF1}
\ee
where the vector notation $n^a = \vec{n}$ is used for simplicity. In this case, we will be interested in the Epstein-Hurwitz series
\be
E_F = \sum_{\vec{n} \in \mathbb{Z}^d} \frac{1}{|\vec{n} + a|^2} \, ,
\ee
when $a \not= 0$. Here again, we will modify the sum to include a regulator $\alpha^2$, which trunctates the modes above $N \sim \alpha^{-1}$ out of the sum. In the present case, the expression of interest will be
\be
\sum_{\vec{n} \in \mathbb{Z}^d} \frac{e^{-\alpha^2 |\vec{n}+a|^2}}{|\vec{n}+a|^2} \, ,
\ee
which reduces to $E_F$ when $\alpha^2$ goes to zero. Making use of equation \ref{eq:int_indentity1}, the sum above can be rewritten as an integral. We obtain
\begin{align}
\sum_{\vec{n} \in \mathbb{Z}^d} \frac{e^{-\alpha^2 |\vec{n}+a|^2}}{|\vec{n}+a|^2} & = \int_0^\infty dt \sum_{\vec{n} \in \mathbb{Z}^d} e^{-(t + \alpha^2) |\vec{n}+a|^2} \\
& = \pi \int_{\alpha^2/\pi}^{\infty} dt \, \theta^d(t|a) \, ,
\label{eq:intF} 
\end{align}
where we defined the function 
\be
\theta(t|a) = \sum_{n = -\infty}^{\infty} e^{-\pi t (n+a)^2} \, .
\ee
Just like $\theta(t)$, the function above can be approximated as $\theta(t|a) \sim t^{-1/2}$ when $t \rightarrow 0$, so the integrand in equation \ref{eq:intF} diverges in the limit when the regulator $\alpha^2$ goes to zero.  To treat this divergence, we will rewrite the divergent part of the integral by making use of the property
\be
\theta(t|a) = \frac{e^{-\pi a^2 t}}{t^{1/2}} \theta(1/t|iat)
\ee
which can be derived using Poisson's resummation formula. In this case, the divergent part of the integral can be written as
\be
\int_{\alpha^2/\pi}^{1} dt \, \theta^d(t|a) = \int_{1}^{\pi/\alpha^2} dt \, t^{d/2-1} \left( e^{-\pi d a^2 / t}\theta^d(t|ia/t) - 1 \right) - \frac{2}{d-2} + \frac{2}{d-2} \left( \frac{\pi}{\alpha^2} \right)^{d/2-1} \, .
\ee
The integral on the right-hand side of the expression above is convergent for all values of $d$. Hence, when $d/2 > 1$, the only divergent piece comes from the last term which is inversely proportional to $\alpha^2$. Piecing everything together and letting $\alpha^2$ go to zero, we obtain
\be 
E_F = \pi \left( \int_{1}^{\infty} dt \, \theta^d(t|a) + \int_{1}^{\infty} dt \, t^{d/2-1} \left( e^{-\pi d a^2 / t}\theta^d(t|ia/t) - 1 \right) - \frac{2}{d-2} + \frac{2}{d-2} \left( \frac{\pi}{\alpha^2} \right)^{d/2-1} \right) \, .
\ee
Letting $d = 6$ and $a = 1/2$, we can finally evaluate $S_{F_1}$ by substituting $E_F$ in equation \ref{eq:SF1}. We obtain
\be 
S_{F_1} = \frac{1}{4\pi} \left( \int_{1}^{\infty} dt \, \theta^6(t|1/2) + \int_{1}^{\infty} dt \, t^{2} \left( e^{-\frac{3\pi}{2t}}\theta^6(t|i(2t)^{-1}) - 1 \right) - \frac{1}{2} + \frac{1}{2} \left( \frac{\pi}{\alpha^2} \right)^{2} \right) \, .
\ee
As we can see, the divergent piece in $S_{F1}$ is the same one that we obtained for $S_{B1}$. This is decause in the $t \rightarrow 0$ limit, the theta function in the integrand (\ref{eq:intF}) behaves as $\theta(t|a) \sim t^{-1/2}$ independantly of $a$. Consequently, integrating $\theta(t|a)^d$ in the viscinity of $t \rightarrow 0$ yields the same divergent piece regardless of if $a$ takes the value zero (in the bosonic case) or $1/2$ (in the fermionic case). This means that substracting $S_{B_1}$ from $S_{F_1}$ should give a finite value, which can be otained by carrying out each integrals in $S_{B_1}$ and $S_{F_1}$. Carrying out the integrals numerically, we obtain
\be
S_{F1} - S_{B_1} = 0.0397887 \, .
\ee

\end{document}